\documentclass[12pt]{article}
\usepackage{amsmath}
\usepackage{amssymb}
\usepackage{epsfig}
\usepackage{euscript}
\usepackage{fancybox}
\usepackage{color}
\def\mathswitchr#1{\relax\ifmmode{\mathrm{#1}}\else$\mathrm{#1}$\fi}

\newcommand {\pslash}{\hbox{$\not\hbox{\kern-2.3pt $p$}$}}

\usepackage{cite}

\usepackage{epic}
\def\alf1{ {\alpha\over\pi} }
\begin{document}
%\input{feynman} 
%=======================================================================
\begin{titlepage}
\begin{flushright}
%{\bf MPI-PhT-2002-08}\\
 {\bf BU-HEPP-04-06 }\\
{\bf Nov, 2004}\\
\end{flushright}
%\vspace{0.05cm}
 
\begin{center}
{\Large Black Holes and Massive Elementary Particles in Resummed Quantum Gravity$^{\dagger}$
}
\end{center}

\vspace{2mm}
\begin{center}
%%  {\bf   S. Jadach$^{a,b}$ and B.F.L. Ward$^{c,d}$}
{\bf   B.F.L. Ward}\\
\vspace{2mm}
%{\em $^a$CERN, Theory Division, CH-1211 Geneva 23, Switzerland,}\\
%{\em $^b$Institute of Nuclear Physics,
%        ul. Kawiory 26a, Krak\'ow, Poland,}
%{\em $^c$Werner-Heisenberg-Institut, Max-Planck-Institut fuer Physik,
%Muenchen, Germany,}\\
%{\em $^a$Werner-Heisenberg-Institut, Max-Planck-Institut fuer Physik,
%Muenchen, Germany,}\\
%{\em $^d$Department of Physics and Astronomy,\\
%  The University of Tennessee, Knoxville, Tennessee 37996-1200, USA.}\\
%{\em $^c$SLAC, Stanford University, Stanford, California 94309, USA,}\\
{\em Department of Physics,\\
  Baylor University, Waco, Texas 76798-7316, USA}\\
%{\em and }\\
%{\em Department of Physics and Astronomy,\\
%  The University of Tennessee, Knoxville, Tennessee 37996-1200, USA}\\
%{\em $^c$SLAC, Stanford University, Stanford, California 94309, USA,}\\
\end{center}

%\begin{center}
% {\bf S. Jadach}\\
%{DESY, Theory Division, D-22603 Hamburg, Germany}\\
%   {\em Institute of Nuclear Physics,
%        ul. Kawiory 26a, Krak\'ow, Poland}\\
%   {\em CERN, Theory Division, CH-1211 Geneva 23, Switzerland,}\\
% {\bf W.  P\l{a}czek}\\
%   {\em Institute of Computer Science,
%   Jagellonian University, ul. Nawojki 11, 30-072 Krak\'ow, Poland}\\
%{\em CERN, Theory Division, CH-1211 Geneva 23, Switzerland,}\\
% {\bf M. Skrzypek}\\
%   {\em Institute of Nuclear Physics,
%        ul. Kawiory 26a, Krak\'ow, Poland}\\
%{\em CERN, Theory Division, CH-1211 Geneva 23, Switzerland,}\\
% {\bf B.F.L. Ward}\\
%   {\em Department of Physics and Astronomy,\\
%   The University of Tennessee, Knoxville, Tennessee 37996-1200\\
%   SLAC, Stanford University, Stanford, California 94309}\\
%{\em CERN, Theory Division, CH-1211 Geneva 23, Switzerland,}\\
%{\bf Z. W\c as}\\
%   {\em Institute of Nuclear Physics,
%        ul. Kawiory 26a, Krak\'ow, Poland}\\
%   {\em CERN, Theory Division, CH-1211 Geneva 23, Switzerland}\\
%\end{center}

\vspace{5mm}
\begin{center}
{\bf   Abstract}
\end{center}
Einstein's general theory of relativity poses many problems
to the quantum theory of point particle fields. Among them is the fate of a
massive point particle. Since its rest mass exists entirely
within its Schwarzschild radius, in the classical solutions of
Einstein's theory, the respective system should be a black hole.
We address this issue using exact results in a new approach to 
quantum gravity based upon well-tested resummation methods in
point particle quantum field theory. We
show that the classical conclusion is obviated by quantum loop effects.
We show that our new approach already passes two theoretical checks
with the published literature; for, it reproduces known results 
on the one-loop correction to the graviton self-energy in scaler matter
coupled to Einstein's gravity as analyzed by 't Hooft and Veltman
and it is consistent with the asymptotic safety results of Bonnanno
and Reuter on the behavior of Newton's constant in the deep Euclidean
regime. Indeed, our approach is consistent with the black hole
phenomenology of the latter authors, including their results on the
final state of the Hawking radiation for an originally massive black hole.
Further black hole related phenomenological implications are 
also discussed.\\
\vspace{10mm}
% 
%\vspace{10mm}
\\
\centerline{ Submitted to Nova Science Publishers }
\renewcommand{\baselinestretch}{0.1}
\footnoterule
\noindent
{\footnotesize
\begin{itemize}
\item[${\dagger}$]
Work partly supported 
% the Polish Government
%grants KBN 2P30225206 and 2P03B17210, the Maria Sk\l{}odowska-Curie
%Joint Fund II PAA/DOE-97-316, and
by NATO Grant PST.CLG.980342.
%, and by
%Polish Government grant 5P03B09320.
\end{itemize}
}
%\vspace{0.5cm}
%\centerline{ Submitted to JCAP }
%\begin{flushleft}
%{\bf UTHEP-00-0101}\\
%{\bf Jan, 2000}\\
%\end{flushleft}

\end{titlepage}

%=======================================================================
\def\Kmax{K_{\rm max}}\def\ieps{{i\epsilon}}\def\rQCD{{\rm QCD}}
\renewcommand{\theequation}{\arabic{equation}}
\font\fortssbx=cmssbx10 scaled \magstep2
\renewcommand\thepage{}
%\vfill\eject
\parskip.1truein\parindent=20pt\pagenumbering{arabic}\par
\section{\bf Introduction}\label{intro}\par
Einstein's general theory of relativity has had many 
successful tests~\cite{mtw,sw1}. One of its most direct classical
predictions is that a massive elementary point particle should be
a black hole solution; for, in such a system, the rest mass lies entirely
within the Schwarzschild radius. For the Standard Model~\cite{sm,qcd1}
such a conclusion would be at best problematic, as many of the
elementary particles in the SM have non-zero rest mass and yet
they are able to communicate themselves entirely in the well-tested
SM point-particle field theoretic interactions -- they do not only interact
with the world beyond their Schwarzschild radii by Hawking 
radiation~\cite{hawk}. A key point then is whether or not quantum
loop effects obviate the classical general relativistic conclusion that
a massive point particle is a black hole. It is this question that 
we answer in the following.\par

We base our analysis on pioneering work by Feynman~\cite{rpf1,rpf2}
on the quantum theory of general relativity in which he argued
that Einstein's theory is just another point particle field theory
in which the metric field of space-time undergoes quantum fluctuations
as do all the other point particle fields in the SM. Feynman worked-out
the Feynman rules for the simplest case of a massive scalar field
coupled to Einstein's gravity and we will use his results in what follows.
His formulation of quantum general relativity however was badly behaved
in the deep Euclidean (ultra-violet(UV)) regime and 
it can be characterized as being 
non-renormalizable~\cite{itzb,sw2}.
What we do pedagogically beyond what is done in Refs.~\cite{rpf1,rpf2}
is to re-arrange {\it exactly} the Feynman series 
for quantum general relativity
derived therein using our recent extension of YFS~\cite{yfs,yfs1} resummation methods to non-Abelian gauge theories~\cite{qcdyfs}. The resummed theory
which we derive~\cite{bw1}, which will be seen to have improved UV behavior, 
is therefore not an approximation. We call it resummed quantum
gravity (RQG). Since we do not modify the theory of Einstein or the
quantum mechanics, we arrive at a minimal union of the ideas of
Bohr and Einstein, in contrast to the popular superstring theory\cite{gsw,jp}
or the recently developing loop quantum gravity theory~\cite{lpg1}(LQG),
where the existence of the smallest length parameter, the Planck length,
means that at least Einstein's theory is modified in these two
approaches to finite quantum loop effects in quantum general relativity\footnote{In the superstring theory, it appears that quantum mechanics is also
modified~\cite{gross}.}.\par

The basic physical idea behind our resummation efforts is the following.
In the over-all space-time point of view of Feynman, a point particle of mass
$m$ at a point $x$ and another such particle at $y$  experience an attractive
force $\propto m^2$ due to Newton's law. When this is considered in the deep
Euclidean regime, wherein the effective value of $m^2$ is large and negative,
it translates into a large repulsive interaction against the propagation
of the particle between the respective $x$ and $y$. 
One concludes that the propagation
of the particle should be severely damped in the deep Euclidean regime
in the exact solutions of quantum general relativity. This suggests
that we resum the theory to get better behavior of the Feynman series
in the deep Euclidean regime.\par

We point-out that, in Ref.~\cite{wein1}, Weinberg has noted the four
approaches to the bad UV behavior of quantum general relativity:
\begin{itemize}
\item extended theories of gravitation: supersymmetric theories - 
superstrings; loop quantum gravity, etc.
\item resummation 
\item composite gravitons
\item asymptotic safety: fixed point theory (see Ref.~\cite{laut,reuter2})
%%%lautscher \& reuter,  hep-th/0205062)} 
\end{itemize} 
Here, we are developing a new version of the resummation approach.
We will make contact with the more phenomenological asymptotic
safety approach as realized in Refs.~\cite{laut,reuter2} 
especially as it relates
to black hole physics. Some speculations about the possible relationship
between our new RQG theory and the superstring theory can be found
in Ref.~\cite{bw1,bw2}, wherein we also point-out the complementarity
between our analysis and the large distance analyses in Ref.~\cite{dono1}.\par

More precisely, after reviewing the formulation of Einstein's theory by Feynman
in the next Section, we use YFS resummation methods in Section 3 to 
arrive at a theory of quantum general relativity 
in which the loop corrections are finite. In Section 4, we use these 
finite loop corrections
to address the fate of massive elementary particles from the
standpoint of their being black holes. We also show in this Section  
how our results for the UV behavior of our quantum
loop effects relate to the corresponding results of the
asymptotic safety approach in Refs.~\cite{laut,reuter2}.
We argue that this relationship leads us to the same
black hole physics phenomenology as that found in Refs.~\cite{laut,reuter2}.
Section 5 contains some summary remarks.\par

\section{ Point Particle Field Theoretic Formulation of Einstein's Theory}
 
In Feynman's point particle formulation~\cite{rpf1,rpf2} of Einstein's theory
we start with
the Lagrangian density of the currently observed world
\begin{equation}
%\begin{split}
{\cal L}(x) = -\frac{1}{2\kappa^2}\sqrt{-g} R
            + \sqrt{-g} L^{\cal G}_{SM}(x)
\label{lgwrld}
\end{equation}
where $R$ is the curvature scalar, $-g$ is the
negative of the determinant of the metric of space-time
$g_{\mu\nu}$, $\kappa=\sqrt{8\pi G_N}\equiv 
\sqrt{8\pi/M_{Pl}^2}$, where $G_N$ is Newton's constant,
and the SM Lagrangian density, which is well-known
( see for example, Ref.~\cite{sm,qcd1,barpass} ) when invariance 
under local Poincare symmetry is not required,
is here represented by $L^{\cal G}_{SM}(x)$ which is readily obtained
from the familiar SM Lagrangian density as described in
Refs.~\cite{bw1,bw2} as follows:
since $\partial_\mu\phi(x)$ is already generally
covariant for any scalar field $\phi$ and since the only derivatives of the
vector fields in the SM Lagrangian density occur in their
curls, $\partial_\mu A^J_\nu(x)-\partial_\nu A^J_\mu(x)$, which are
also already generally covariant, we only need
to give a rule for making the fermionic terms in 
usual SM Lagrangian density generally covariant. For this,
we introduce a differentiable structure with $\{\xi^a(x)\}$ as
locally inertial coordinates and an attendant
vierbein field $e^a_\mu\equiv\partial\xi^a/\partial x^\mu$ 
with indices that carry 
the vector representation for the flat locally inertial space, $a$, and for the
manifold of space-time, $\mu$, with the identification of the space-time
base manifold metric as
$g_{\mu\nu}=e^a_\mu e_{a\nu}$ where the flat locally inertial 
space indices are to be
raised and lowered with Minkowski's metric $\eta_{ab}$ as usual. 
Associating the usual Dirac gamma
matrices $\{\gamma_a\}$ with the flat locally inertial space at x, we define
base manifold Dirac gamma matrices by $\Gamma_\mu(x)=e^a_\mu(x)\gamma_a$.
Then the spin connection, $\omega_{\mu b}^a=-\frac{1}{2}e^{a\nu}\left(
\partial_\mu e^b_\nu-\partial_\nu e^b_\mu\right)+\frac{1}{2}e^{b\nu}\left(
\partial_\mu e^a_\nu-\partial_\nu e^b_\mu\right)
+\frac{1}{2}e^{a\rho}e^{b\sigma}\left(\partial_\rho e_{c\sigma}-\partial_\sigma e_{c\rho}\right)e^c_\mu$ when there is no torsion, allows us to 
identify the generally covariant
Dirac operator for the SM fields by the substitution
${i\not{\partial}} \rightarrow i\Gamma(x)^\mu\left(\partial_\mu +\frac{1}{2}{\omega_{\mu b}}^a{\Sigma^b}_a\right)$, where we have ${\Sigma^b}_a=\frac{1}{4}\left[\gamma^b,\gamma_a\right]$
everywhere in the SM Lagrangian density. This will generate $L^{\cal G}_{SM}(x)$ from the usual SM Lagrangian density $L_{SM}(x)$ as it is
given in Refs.~\cite{sm,qcd1,barpass}, for example.\par

The fundamental issue we address here is the fate of the massive
point particles in the SM. We do not expect the respective 
spin representation to be 
crucial to the conclusions we reach so we follow Feynman and
replace (\ref{lgwrld}) with the simplest example of our problem, the
Lagrangian of a massive scalar field
coupled to Einstein's gravity, where we have in mind the physical
Higgs field of the SM, whose rest mass is known to be greater
than 114 GeV with a 95\% CL~\cite{lewwg}:
\begin{equation}
\begin{split}
{\cal L}(x) &= -\frac{1}{2\kappa^2} R \sqrt{-g}
            + \frac{1}{2}\left(g^{\mu\nu}\partial_\mu\varphi\partial_\nu\varphi - m_o^2\varphi^2\right)\sqrt{-g}\\
            &= \quad \frac{1}{2}\left\{ h^{\mu\nu,\lambda}\bar h_{\mu\nu,\lambda} - 2\eta^{\mu\mu'}\eta^{\lambda\lambda'}\bar{h}_{\mu_\lambda,\lambda'}\eta^{\sigma\sigma'}\bar{h}_{\mu'\sigma,\sigma'} \right\}\\
            & \qquad + \frac{1}{2}\left\{\varphi_{,\mu}\varphi^{,\mu}-m_o^2\varphi^2 \right\} -\kappa {h}^{\mu\nu}\left[\overline{\varphi_{,\mu}\varphi_{,\nu}}+\frac{1}{2}m_o^2\varphi^2\eta_{\mu\nu}\right]\\
            & \quad - \kappa^2 \left[ \frac{1}{2}h_{\lambda\rho}\bar{h}^{\rho\lambda}\left( \varphi_{,\mu}\varphi^{,\mu} - m_o^2\varphi^2 \right) - 2\eta_{\rho\rho'}h^{\mu\rho}\bar{h}^{\rho'\nu}\varphi_{,\mu}\varphi_{,\nu}\right] + \cdots \\
\end{split}
\label{eq1}
\end{equation}
Here, 
%$\varphi(x)$ is our representative scalar field for matter,
$\varphi(x)_{,\mu}\equiv \partial_\mu\varphi(x)$,
and $g_{\mu\nu}(x)=\eta_{\mu\nu}+2\kappa h_{\mu\nu}(x)$ 
where we follow Feynman and expand about Minkowski space
so that $\eta_{\mu\nu}=diag\{1,-1,-1,-1\}$. 
Following Feynman, we have introduced the notation
$\bar y_{\mu\nu}\equiv \frac{1}{2}\left(y_{\mu\nu}+y_{\nu\mu}-\eta_{\mu\nu}{y_\rho}^\rho\right)$ for any tensor $y_{\mu\nu}$\footnote{Our conventions for raising and lowering indices in the 
second line of (\ref{eq1}) are the same as those
in Ref.~\cite{rpf2}.}. 
Thus, $m_o$ is the bare mass of our free Higgs field and we set the small
tentatively observed~\cite{cosm1} value of the cosmological constant
to zero so that our quantum graviton has zero rest mass.
%Here, our normalizations are such that $\kappa=\sqrt{8\pi G_N}$
%where $G_N$ is Newton's constant.
The Feynman rules for (\ref{eq1}) have been essentially worked out by 
Feynman~\cite{rpf1,rpf2}, including the rule for the famous
Feynman-Faddeev-Popov~\cite{rpf1,ffp1} ghost contribution that must be added to
it to achieve a unitary theory with the fixing of the gauge
( we use the gauge of Feynman in Ref.~\cite{rpf1}, 
$\partial^\mu \bar h_{\nu\mu}=0$ ), 
so we do not repeat this 
material here.\par

\begin{figure}
\begin{center}
\epsfig{file=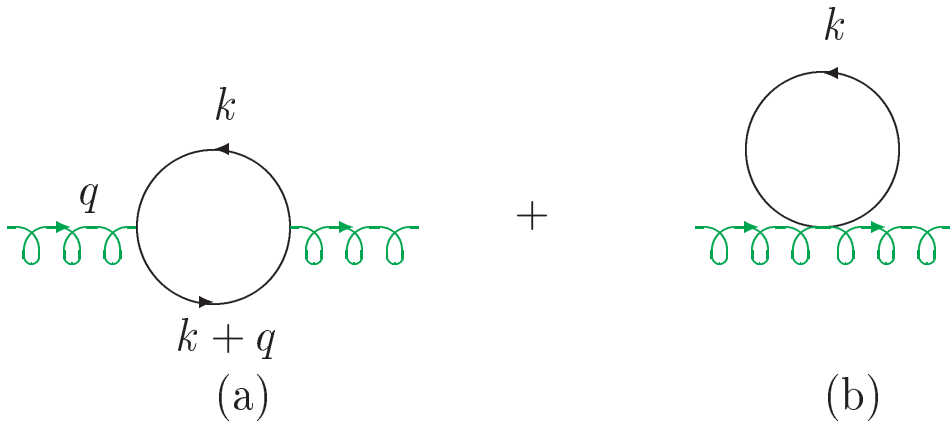,width=140mm}
\end{center}
\caption{\baselineskip=7mm     The scalar one-loop contribution to the
graviton propagator. $q$ is the 4-momentum of the graviton.}
\label{fig1}
\end{figure}

We wish to use the quantum loop corrections predicted by
(\ref{eq1}) to study the black hole character of our massive
elementary particle $\varphi$. The original Feynman
series derived in Refs.~\cite{rpf1,rpf2} is badly UV divergent.
For example, the graphs in Fig.~\ref{fig1} are superficially quarticly
divergent and, since they give us the scalar one-loop contribution to the
graviton propagator, they will lead to a logarithmic UV divergence
in the coefficient of $q^4$ in the respective 1PI 2-point function,
a divergence which can not be removed by any amount of field and 
mass renormalization,
i.e., these graphs already exhibit the non-renormalizability of the
original Feynman series derived in Ref.~\cite{rpf1,rpf2}.
In the next Section, with an eye toward our black hole physics analysis,
we improve on it here by using the extension of the methods
of Yennie, Frautschi and Suura~\cite{yfs} to non-Abelian gauge theories
as we have developed this extension in Ref.~\cite{qcdyfs}.\par

\section{Resummed Quantum Gravity}

In this section,
we will YFS resum the propagators in the theory in (\ref{eq1}):
from the YFS formula (see eq.(5.18) in Ref.~\cite{yfs})
%\begin{equation}
%\Sigma_F(p)=e^{\alpha B''_\gamma}\left[{\Sigma'}_F(p)-S^{-1}_F(p)\right]+
%S^{-1}_F(p),
%\end{equation}
%which implies
\begin{equation}
iS'_F(p) = \frac{ie^{-\alpha B''_\gamma}}{S^{-1}_F(p)-{\Sigma'}_F(p)},
\label{yfsa}
\end{equation}
%for
%\begin{equation}
%{\Sigma'}_F(p)=\sum_{n=1}^{\infty}{\Sigma'}_{Fn},
%\end{equation}
where ${\Sigma'}_F(p)$ is the sum of the YFS loop residuals,
we need to find for quantum gravity the analogue of
\begin{equation}
 \alpha B''_\gamma = \int d^4\ell\frac{S''(k,k,\ell)}{\ell^2-\lambda^2+i\epsilon}
\label{virt1}
\end{equation}
where $\lambda$ is the IR cut-off and {\small 
\begin{eqnarray}
S''(k,k,\ell) = \frac{-i8\alpha}{(2\pi)^3}\frac{kk'}{(\ell^2-2\ell k+\Delta+i\epsilon)}\nonumber\\
\frac{1}{(\ell^2-2\ell k'+\Delta'+i\epsilon)}{\Big|}_{k=k'},
\end{eqnarray}}
for
$\Delta =k^2 - m^2$,~$\Delta' ={k'}^2 - m^2$.
To this end, note also 
{\small
\begin{eqnarray}
\alpha B''_\gamma &= \int \frac{d^4\ell}{(2\pi)^4}\frac{-i\eta^{\mu\nu}}{(\ell^2-\lambda^2+i\epsilon)}\frac{-ie(2ik_\mu)}{(\ell^2-2\ell k+\Delta+i\epsilon)}\nonumber\\
&\frac{-ie(2ik'_\nu)}{(\ell^2-2\ell k'+\Delta'+i\epsilon)}{\Big|}_{k=k'},
\label{virt2}
\end{eqnarray}}
where $\Delta =k^2 - m^2$,~$\Delta' ={k'}^2 - m^2$ and
$\lambda$ is the IR cut-off. With the 
identifications~\cite{sw2} of the conserved 
graviton charges via 
$e \rightarrow \kappa k_\rho$ for soft emission from $k$
we get the analogue ,$-B''_g(k)$, of $\alpha B''_\gamma$
by replacing the $\gamma$ propagator in (\ref{virt2}) by the graviton 
propagator, 
$$\frac{i\frac{1}{2}(\eta^{\mu\nu}\eta^{\bar\mu\bar\nu}+
\eta^{\mu\bar\nu}\eta^{\bar\mu\nu}-\eta^{\mu\bar\mu}\eta^{\nu\bar\nu})}{\ell^2-\lambda^2+i\epsilon}$$, 
and by replacing the QED charges by the corresponding gravity charges $\kappa k_{\bar\mu},~\kappa k'_{\bar\nu}$. 
This yields~\cite{bw1} for our scalar propagator
\begin{equation}
i\Delta'_F(k)|_{Resummed} =  \frac{ie^{B''_g(k)}}{(k^2-m^2-\Sigma'_s+i\epsilon)}.
\label{resum}
\end{equation}
where
\begin{eqnarray} 
B''_g(k)= -2i\kappa^2k^4\frac{\int d^4\ell}{16\pi^4}\frac{1}{\ell^2-\lambda^2+i\epsilon}\nonumber\\
\frac{1}{(\ell^2+2\ell k+\Delta +i\epsilon)^2}
\label{yfs1} 
\end{eqnarray}
so that $B''_g(k) = \frac{\kappa^2|k^2|}{8\pi^2}\ln\left(\frac{m^2}{m^2+|k^2|}\right)$ in the deep Euclidean regime. If $m$ vanishes, using the usual $-\mu^2$ normalization 
point we get
$B''_g(k)=\frac{\kappa^2|k^2|}{8\pi^2}
\ln\left(\frac{\mu^2}{|k^2|}\right)$. In both cases the
resummed propagator falls faster than any power of $|k^2|$!
%\begin{equation}
%i\Delta'_F(k)|_{Resummed} =  \frac{ie^{B''_g(k)}}{(k^2-m^2-\Sigma'_s+i\epsilon)}.
%\label{resum}
%\end{equation}
This is the basic result.
Note that
$\Sigma'_s$ starts in ${\cal O}(\kappa^2)$, so we may drop it in
calculating the one-loop effects which are at the heart of our black hole
physics analysis. 
%Further,
%explicit evaluation gives, for the deep UV regime,
%\begin{equation}
%B''_g(k) = \frac{\kappa^2|k^2|}{8\pi^2}\ln\left(\frac{m^2}{m^2+|k^2|}\right).
%\label{deep}
%\end{equation}
%The resummed propagator falls faster than any power of $|k^2|$!
%If $m$ vanishes, using the usual $-\mu^2$ normalization 
%point we get
%$B''_g(k)=\frac{\kappa^2|k^2|}{8\pi^2}
%\ln\left(\frac{\mu^2}{|k^2|}\right)$
%which again vanishes faster than any power of $|k^2|$! 
Our result that the resummed propagator falls faster than any power
of $|k^2|$ means that one-loop corrections are finite! 
Indeed, all quantum gravity loops are UV finite and the all orders
proof, as well as the explicit finiteness of ${\Sigma'_s}$
at one-loop, 
is given in Refs.~\cite{bw1}.\par
\section{Massive Elementary Particles and Black Hole Physics in RQG}

The one-loop corrections to Newton's law
implied by the diagrams in Fig.~\ref{fig1} are central to our discussion, as
they directly impact our black hole physics issues.
Using the YFS resummed propagators in Fig.~\ref{fig1}, as we have
shown in Refs.~\cite{bw1,bw2} we get
the Newtonian potential
$\Phi_{N}(r)= -\frac{G_N M_1M_2}{r}(1-e^{-ar})$
%%\end{equation}
%where $a=1/\sqrt{-\frac{1}{2}\Sigma^{T(2)}}\simeq 3.96 M_{Pl}$
where~\cite{bw1,bw2} $a\simeq 3.96 M_{Pl}$
when for definiteness we set $m\cong 120$GeV.
%We note that
%%\begin{equation}
%$c_2 \cong \ln\frac{1}{\lambda_c}-\ln\ln\frac{1}{\lambda_c}-\frac{\ln\ln\frac{1}{\lambda_c}}{\ln\frac{1}{\lambda_c}-\ln\ln\frac{1}{\lambda_c}}-\frac{11}{6}$.
%%\label{anal1}
%%\end{equation}
%%and we used this result to check our numerical result for $c_2$.
%Without resummation, $\lambda_c=0$
%and $c_2$ is infinite.
%% and, as this is
%%the coefficient of $q^4$ in the inverse propagator, 
%%{\bf no renormalization of the field and/or of the mass could remove
%%such an infinity}. In our new approach, 
%%this infinity is absent.\par
%Our gauge invariant result for $\Sigma^{T(2)}$
%% in (\ref{sigma})
Our gauge invariant analysis
can be shown~\cite{bw1} to be consistent
with the one-loop analysis of QG in Ref.~\cite{thvelt1}\footnote{Our deep Euclidean studies are complementary 
to the low energy studies
of Ref.~\cite{dono1}.}.
%%CROSS CHECK WITH 't HOOFT AND VELTMAN, {\Color{Red}Ann. Inst. Henri Poincare {\bf XX}, 69 (1974)}, {\Color{Black}WHERE THE COMPLETE
%%RESULT OF THE ONE-LOOP DIVERGENCES OF OUR SCALAR FIELD COUPLED
%%TO EINSTEIN'S GRAVITY HAVE BEEN COMPUTED}. 
%%\newpage

%%Sub-Planck scale physics is accessible to point particle field theory
%%so that current superstring theories may be
%%phenomenological models
%%for a more fundamental theory (TUT=The Ultimate Theory) just as the old string theory~\cite{schw1} is such a model for QCD. Other types of correspondences
%%are not excluded here~\cite{superichep04}.
%%Our deep Euclidean studies are complementary 
%%to the low energy studies
%%of Ref.~\cite{dono1}. 
%%The effective cut-off which we generate dynamically
%%is at $M_{Pl}$ so that renormalizable quantum field theory (QFT)  
%%below $M_{Pl}$
%%is unaffected. Some non-renormalizable QFT's are given new 
%%life here -- they may have other problems, however.

%\section{Massive Elementary Particles and Black Holes}

%In the SM, there are 
%now believed to be three massive neutrinos~\cite{neut},
%with masses that we estimate at $\sim 3$ eV, and there are 
%the remaining members
%of the known three generations of Dirac fermions 
%$\{e,\mu,\tau,u,d,s,c,b,t\}$. 
With reasonable estimates and measurements 
~\cite{neut,pdg2002,bw1} of the SM particle 
masses, including the various bosons,
%$m_e\cong 0.51$ MeV, $m_\mu \cong 0.106$ GeV, $m_\tau \cong 1.78$ GeV,
%$m_u \cong 5.1$ MeV, $m_d \cong 8.9$ MeV, $m_s \cong 0.17$ GeV,
%$m_c \cong 1.3$ GeV, $m_b \cong 4.5$ GeV and $m_t \cong 174$ GeV,
%as well as the massive vector bosons $W^{\pm},~z$, with masses
%$M_W\cong 80.4$ GeV,~$M_Z\cong 91.19$ GeV.  
%
%To get a better estimate of the size of $c_2$ we 
%use the general spin independence
%of the graviton coupling to matter at {\it relatively low} momentum transfers.
%We count each Dirac fermion as 4 degrees of freedom,
%each massive vector boson as 3 degrees of freedom and remember that
%each quark has three colors. 
the corresponding results for the analogs of the
diagrams in Fig. 1 imply~\cite{bw1}
%$c_2$ in Ref.~\cite{bw2} for each
%SM massive degree of freedom implies approximately
%\begin{equation}
%$c_{2,eff} \cong 9.26\times 10^3$
%\label{ceff} 
%\end{equation}
%so 
that in the SM
%\begin{equation}
$a_{eff} \cong 0.349 M_{Pl}$ .
%\label{aeff} 
%\end{equation}
To make direct contact with black hole physics, 
note that, if $r_S$ is the Schwarzschild radius,
for $r\rightarrow r_S$, $a_{eff}r \ll 1$ so 
that $|2\Phi_{N}(r)|_{m_1=m}/m_2|\ll 1$. This means that
$g_{00}\cong 1+2\Phi_{N}(r)|_{m_1=m}/m_2$ remains 
positive as we pass through the
Schwarzschild radius. 
It can be shown~\cite{bw1} that this 
positivity holds to $r=0$. Similarly, $g_{rr}$ remains negative
through $r_S$ down to $r=0$~\cite{bw1}. 
To get these results,
note that in the relevant regime for r, the smallness of
the quantum corrected newton potential means that we can use the
linearized Einstein equations for a small spherically symmetric
static source $\rho(r)$ which generates $\phi_{newton}(r)|_{m_1=m}/m_2$
via the standard Poisson's equation. 
the usual result (see Refs.~\cite{mtw,abs}) for the
respective metric solution then gives 
$g_{00}\cong 1+2\phi_{newton}(r)|_{m_1=m}/m_2$ and
$g_{rr}\cong -1+2\phi_{newton}(r)|_{m_1=m}/m_2$ 
which remain
respectively time-like and space-like
to $r=0$.\par
In resummed QG, a massive SM point particle is not a black hole.\par

The value of $a_{eff}$ given here is incomplete,
as there may be as yet unknown massive particles 
beyond those already
discovered -- these would only decrease $a_{eff}$.
%Including more particles in the computation of
%$a_{eff}$ would make it smaller and hence would not change the
%conclusions of our analysis. 
For example, in the minimal 
supersymmetric Standard
Model we expect approximately that 
$a_{eff}\rightarrow \frac{1}{\sqrt{2}}a_{eff}$.

One can also use the results for the
complete one-loop UV divergent corrections of Ref.~\cite{thvelt1}
to see that the remaining 
interactions at one-loop order not discussed here
(vertex corrections, pure gravity self-energy corrections, etc. )
also do not increase the value of $a_{eff}$ --
$a_{eff}$
is a parameter which is bounded from above by the estimates we give
here. 
%Indeed, from Ref.~\cite{thvelt1}, we can estimate that 
%the remaining one-loop gravity interactions contribute to $a_{eff}$
%xxxxxxxxx
%%%START HERE.
In general, we expect that the precise value of $a_{eff}$  should be determined from cosmological and/or other
considerations. Such implications will be taken up elsewhere.\par

Our results imply the running Newton constant
$G_N(k)=G_N/(1+\frac{k^2}{a_{eff}^2})$
which is 
fixed point behavior for 
$k^2\rightarrow \infty$,
in agreement with the phenomenological asymptotic safety approach of
Ref.~\cite{reuter2}.
Our result that an elementary particle has no horizon
also agrees with the result in Ref.~\cite{reuter2} that a black hole
with a mass less than
 $M_{cr}\sim M_{Pl}$
has no horizon. The basic physics is the same: $G_N(k)$ vanishes for $k^2\rightarrow \infty$.

Because our value of the coefficient 
of $k^2$ in the denominator of $G_N(k)$
agrees with that found by Ref.~\cite{reuter2}, 
if we use their prescription for the
relationship between $k$ and $r$
in the regime where the lapse function
vanishes,
we get the same Hawking radiation phenomenology as
they do: a very massive black hole evaporates until it reaches a mass
$M_{cr}\sim M_{pl}$
at which the Bekenstein-Hawking temperature vanishes, 
leaving a Planck
scale remnant.\par

We can carry this argument further as follows.
In Ref.~\cite{reuter2} there is some uncertainty about 
the precise value of $M_{cr}$; for, this value depends on a parameter
$\gamma$ which varies bewteen $0$ and $9/2$.
However, independent of the value of $\gamma$,
the result that the value of $M_{cr}$ is of the order of $M_{Pl}$
means that the quantum corrections to the Newtonian potential
found above must be taken into account in the analysis of
Ref.~\cite{reuter2} when the mass of the black hole approaches
this critical value in the Hawking process. When we do this,
we find that the horizon is obviated~\cite{bw3}. This would mean that
eventually the entire mass of the originally massive black hole
would become accessible to the rest of the universe, which is consistent
with Hawking's recent conclusions~\cite{hawk2} on the very important subject of
information loss in black hole physics.\par

\section{Conclusions}

YFS resummation renders quantum gravity finite
so that quantum loop corrections are now cut off dynamically.
We call the resultant new approach to quantum gravity 
resummed quantum gravity (RQG),
which represents a minimal union of the ideas of Bohr and Einstein.
Physics below the Planck scale is accessible to point particle
quantum field theory. Early universe studies may be able to test 
some of our predictions~\footnote{ This is controversial~\cite{kolb} and is
under investigation.} and we point-out that a theoretical cross check
with the analysis of Ref.~\cite{thvelt1} has been done.
In the discussion presented here, we focused on some consequences
of RQG for black hole physics.\par
%    renormalizable qft below $m_{pl}$ unaffected
%  \item
% {\color{green} new life to {\color{red}some} nonrenormalizable qft's: {\color{red} they may have other problems}
%We have achieved a minimal union of Bohr's and Einstein's ideas.
We have shown
%that we are consistent with
%the result in Ref.~\cite{thvelt1} on the one-loop structure
%of quantum gravity and 
that, contrary to classical expectations,
a massive elementary SM particle is not a black hole in resummed quantum gravity. Our results are also consistent with the asymptotic safety analysis
in Ref.~\cite{reuter2} that a black hole of mass less than a critical
mass $\sim M_{Pl}$ does not have a horizon in quantum gravity
and that the final state of the Hawking radiation of a massive black hole
is a Planck scale remnant, which our recent work now shows becomes
accessible to the rest of the universe due to our finite quantum 
loop corrections as well -- a result which agrees with Hawking's
recent results~\cite{hawk2} on information loss in black hole physics.
Further checks are under investigation.\par

\section*{Acknowledgements}

We thank Profs. S. Bethke and L. Stodolsky for the support and kind
hospitality of the MPI, Munich, while a part of this work was
completed. We thank Prof. S. Jadach for useful discussions.

\end{document}